\documentclass[pra,aps,twocolumn,groupaddress]{revtex4}
\usepackage{amsmath,amssymb,graphicx,verbatim,color,braket}
\DeclareMathAlphabet{\mathpzc}{OT1}{pzc}{m}{it}

\begin{document}

\title{Collective excitations using a room-temperature gas of 3-level atoms in a cavity}

\author{G.~T.~Hickman}
\affiliation{Department of Physics, University of Maryland Baltimore County, Baltimore, MD 21250, USA}

\begin{abstract}
We theoretically investigate an ensemble of three-level room-temperature atoms in a two-mode optical cavity, focusing on the case of counterpropagating light fields. We find that in the limit of large detunings the problem admits relatively simple and intuitive solutions. When a classical field drives one of the transitions, the system becomes equivalent to an ensemble of two-level atoms with suppressed Doppler broadening. In contrast, when the same transition is driven with a single photon, the system collectively behaves like a single two-level atom. This latter result is particularly relevant in the context of quantum nonlinear optics.
\end{abstract}

\maketitle

\section{Introduction}\label{S1}

There is great interest in developing systems for all-optical processing of quantum information \cite{Kok10, Reiserer15}. Experiments performed in the strong coupling regime of cavity quantum electrodynamics (QED) have become increasingly important in this context \cite{Berman94}. This regime has been reached using a wide variety of systems, including trapped single atoms \cite{Tiecke14, Reiserer15, Volz14, Reiserer14}, solid-state quantum emitters \cite{Reithmaier04, Fushman08}, artificial atoms \cite{Wallraff04,Hoi13}, and cold atomic clouds \cite{Beck15, Tiarks16}. The use of warm atomic ensembles has also been considered \cite{Venkataraman13, Borregaard16, Diniz11, Hickman15}, but has faced complications because of effects related to the thermal motion of the atoms.

Compared with single trapped atoms and cold atomic clouds, the theoretical description of warm atomic ensembles interacting with quantized fields is complicated by the presence of Doppler broadening. The resonant frequencies of the atoms, rather than having a single fixed value, are randomly distributed according to the Doppler widths of the relevant transitions. In some circumstances it may be sufficient to first solve the system assuming a fixed atomic frequency, then average the results across the appropriate distribution \cite{Schernthanner94}. However this is often not the case, and in general it is necessary to solve the system self-consistently by taking into account the contributions of all Doppler velocity groups at once \cite{Kurucz11}.

As a result, it has so far been an open question whether cavity QED with room-temperature atomic vapors can be used for applications in quantum nonlinear optics. Photons in a vapor-filled cavity couple to collective excitations of the entire atomic ensemble. Consequently, Doppler broadening of the various velocity groups leads to fast decoherence and can prevent an otherwise useful system from reaching the strong coupling regime.

Here we show that the problem of Doppler broadening in cavity QED can be overcome by taking advantage of Doppler-free two-photon transitions. We use the well-known mathematical technique of adiabatic elimination to derive a simplified effective Hamiltonian for an ensemble of three-level Doppler-broadened atoms in a doubly-resonant cavity. The resulting Hamiltonian is valid in the limit of large detuning from the atomic intermediate state, and describes an ensemble of two-level atoms in which the atoms undergo only two-photon transitions \cite{Gardiner00, Gerry90, Gerry92,  Alexanian95}. We further examine the unitary dynamics of this system for various ways of driving the lower atomic transition. Considering counterpropagating light fields we find that if a classical field is used, the system is equivalent to an ensemble of two-level atoms undergoing single-photon transitions with suppressed Doppler broadening. In the case of a single photon driving field, the ensemble collectively behaves like a single two-level atom. This effective atom absorbs at most one photon from the cavity and then saturates, leading to a nonlinear Rabi frequency that increases with the square root of the number of cavity photons. In our approximation, however, the coupling rates for the effective two-level atoms in both of these cases are always much smaller than what would be expected from a single trapped atom optimally placed within the cavity mode.

The process of adiabatic elimination has been described elsewhere \cite{Gardiner00}, however for the sake of clarity we begin by presenting a complete derivation in Section \ref{S2}. We describe the adiabatic elimination of the intermediate atomic states in the limit of large detunings, and derive an effective Hamiltonian governing the evolution of the system. Section \ref{S3} applies this new Hamiltonian to the cases with a classical field and a single-photon field driving the first atomic transition. Discussion and conclusions are provided in Section \ref{S4}.
\begin{figure}
  \includegraphics[width=.6\columnwidth]{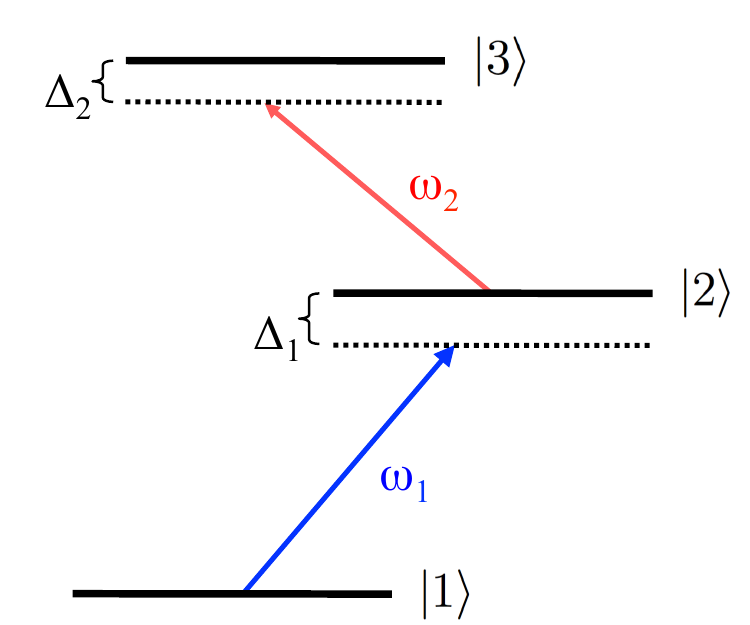}
  \caption{Energy level diagram for a three-level atom in the ladder-type configuration.}\label{fig:ElevelsLadder}
\end{figure}


\section{Adiabatic elimination of the first excited states} \label{S2}

We consider an ensemble of N three-level atoms coupled to a two-mode optical cavity with resonant frequencies $\omega_1$ and $\omega_2$. The energy of level $i$ for atom $j$ is given by $\hbar \omega_{i,j}$. From the perspective of the cavity, the energies $\omega_{i,j}$ for each atom will in general be different because of Doppler broadening. We assume that the atomic levels lie in a ladder-type configuration as shown in Fig. \ref{fig:ElevelsLadder}, though similar results can easily be obtained using the $\Lambda$-type diagram in Fig. \ref{fig:ElevelsLambda}, as discussed in Appendix \ref{SA2}.

The Hamiltonian for this system in the rotating wave approximation is
\begin{equation} \begin{aligned}
\hat{\mathpzc{H}}= &\sum_{j=1}^{N} \sum_{i=1}^{3} \left( \hbar \omega_{i,j} \hat{\sigma}_{ii,j} \right) + \hbar \omega_1 \hat{a}_1^{\mbox{\footnotesize \dag}} \hat{a}_1 + \hbar \omega_2 \hat{a}_2^{\mbox{\footnotesize \dag}} \hat{a}_2 + \\ 
+& \hbar \sum_{j=1}^{N} \bigg[ g_{12,j} \left(\hat{a}_1 \hat{\sigma}_{21,j} + \hat{a}_1^{\mbox{\footnotesize \dag}}\hat{\sigma}_{12,j} \right) + \\ +& g_{23,j} \left( \hat{a}_2 \hat{\sigma}_{32,j} + \hat{a}_2^{\mbox{\footnotesize \dag}} \hat{\sigma}_{23,j} \right) \bigg] ,
\end{aligned}\end{equation}
where $\hat{a}_1^{\mbox{\footnotesize \dag}}$, $\hat{a}_1$, $\hat{a}_2^{\mbox{\footnotesize \dag}}$, and $\hat{a}_2$ are the raising and lowering operators for cavity fields 1 and 2, and the $\hat{\sigma}_{ii,j}$'s are the probabilities for atom $j$ to be found in state $i$. The operators $\hat{\sigma}_{21,j}$ and $\hat{\sigma}_{12,j}$ are the raising and lowering operators respectively for the $\ket{1}$ to $\ket{2}$ transition of atom $j$, and likewise $\hat{\sigma}_{32,j}$ and $\hat{\sigma}_{23,j}$ for the $\ket{2}$ to $\ket{3}$ transition. We assume without loss of generality that the atom-field coupling constants for atom $j$, given by $g_{12,j}$ and $g_{23,j}$, are real and positive. Because of the random positions of the atoms within the cavity mode, these constants are in general different from atom to atom.

We begin by writing the equations of motion in the interaction picture:
\begin{equation} \begin{aligned}
\dot{\hat{a}}_1 = &- i \sum_{j=1}^{N} g_{12,j} \hat{\sigma}_{12,j} \\ \dot{\hat{a}}_2 =& - i \sum_{j=1}^{N} g_{23,j} \hat{\sigma}_{23,j} \\
\dot{\hat{\sigma}}_{12,j} = &-i\Delta_{1,j}\hat{\sigma}_{12,j} + ig_{12,j} \hat{a}_1 \left( \hat{\sigma}_{22,j} - \hat{\sigma}_{11,j} \right) - \\ &- i g_{23,j} \hat{a}_2^{\mbox{\footnotesize \dag}} \hat{\sigma}_{13,j} \\
\dot{\hat{\sigma}}_{23,j} = &-i\Delta_{2,j}\hat{\sigma}_{23,j} + ig_{23,j} \hat{a}_2 \left( \hat{\sigma}_{33,j} - \hat{\sigma}_{22,j} \right) + \\ &+ i g_{12,j} \hat{a}_1^{\mbox{\footnotesize \dag}} \hat{\sigma}_{13,j} \\
\dot{\hat{\sigma}}_{13,j} = &-i \left( \Delta_{1,j} + \Delta_{2,j} \right) \hat{\sigma}_{13,j} + i g_{12,j} \hat{a}_1 \hat{\sigma}_{23,j} - \\ &- i g_{23,j} \hat{a}_2 \hat{\sigma}_{12,j} . \\
\end{aligned} \end{equation}
We have defined $\Delta_{1,j} \equiv \omega_{2,j} - \omega_{1,j} - \omega_{1}$, $\Delta_{2,j} = \omega_{3,j} - \omega_{2,j} - \omega_{2}$. This system is difficult to solve exactly, but we can obtain a solution in the limit of large $\Delta_{1,j}$ and $\Delta_{2,j}$ by finding approximate expressions for $\hat{\sigma}_{12,j}$ and $\hat{\sigma}_{23,j}$. First we note that
\begin{equation} \begin{aligned}
\frac{d}{dt} \bigg[ \hat{\sigma}_{12,j} e^{i \Delta_{1,j}t} \bigg] &= \bigg[ ig_{12,j} \hat{a}_1 \left( \hat{\sigma}_{22,j} - \hat{\sigma}_{11,j} \right) - \\ &- i g_{23,j} \hat{a}_2^{\mbox{\footnotesize \dag}} \hat{\sigma}_{13,j} \bigg] e^{i \Delta_{1,j}t} \\
\frac{d}{dt} \bigg[ \hat{\sigma}_{23,j} e^{i \Delta_{2,j}t} \bigg] &= \bigg[ ig_{23,j} \hat{a}_2 \left( \hat{\sigma}_{33,j} - \hat{\sigma}_{22,j} \right) + \\ &+ i g_{12,j} \hat{a}_1^{\mbox{\footnotesize \dag}} \hat{\sigma}_{13,j} \bigg] e^{i \Delta_{2,j}t} . \label{eq:ddtsigma12,23} \\
\end{aligned} \end{equation}
From Eq. \ref{eq:ddtsigma12,23} integration by parts produces the following \cite{Gardiner00}: \\
\begin{widetext}
\begin{equation}\begin{aligned}
\hat{\sigma}_{12,j}\left( t \right) &= \frac{1}{\Delta_{1,j}} \bigg[ g_{12,j} \hat{a}_1\left( \tau \right) \left( \hat{\sigma}_{22,j}\left( \tau \right) - \hat{\sigma}_{11,j}\left( \tau \right) \right) - g_{23,j} \hat{a}_2^{\mbox{\footnotesize \dag}}\left( \tau \right) \hat{\sigma}_{13,j}\left( \tau \right) \bigg] e^{i \Delta_{1,j}\left(\tau - t \right)} \bigg|_{t_0}^t - \\ &- \frac{1}{\Delta_{1,j}} \int_{t_0}^{t} d\tau \bigg[ g_{12,j} \dot{\hat{a}}_1\left( \tau \right) \left( \hat{\sigma}_{22,j}\left( \tau \right) - \hat{\sigma}_{11,j}\left( \tau \right) \right) + g_{12,j} \hat{a}_1\left( \tau \right) \left( \dot{\hat{\sigma}}_{22,j}\left( \tau \right) - \dot{\hat{\sigma}}_{11,j}\left( \tau \right) \right) - \\ & - g_{23,j} \dot{\hat{a}}_2^{\mbox{\footnotesize \dag}}\left( \tau \right) \hat{\sigma}_{13,j}\left( \tau \right) - g_{23,j} \hat{a}_2^{\mbox{\footnotesize \dag}}\left( \tau \right) \dot{\hat{\sigma}}_{13,j}\left( \tau \right) \bigg] e^{i \Delta_{1,j}\left(\tau - t \right)} \label{eq:sigma12} \\
\end{aligned}\end{equation}
\begin{equation}\begin{aligned}
\hat{\sigma}_{23,j}\left( t \right) &= \frac{1}{\Delta_{2,j}} \bigg[ g_{23,j} \hat{a}_2\left( \tau \right) \left( \hat{\sigma}_{33,j}\left( \tau \right) - \hat{\sigma}_{22,j}\left( \tau \right) \right) + g_{12,j} \hat{a}_1^{\mbox{\footnotesize \dag}}\left( \tau \right) \hat{\sigma}_{13,j}\left( \tau \right) \bigg] e^{i \Delta_{2,j}\left(\tau - t \right)} \bigg|_{t_0}^t  - \\ &- \frac{1}{\Delta_{2,j}} \int_{t_0}^{t} d\tau \bigg[ g_{23,j} \dot{\hat{a}}_2\left( \tau \right) \left( \hat{\sigma}_{33,j}\left( \tau \right) - \hat{\sigma}_{22,j}\left( \tau \right) \right) + g_{23,j} \hat{a}_2\left( \tau \right) \left( \dot{\hat{\sigma}}_{33,j}\left( \tau \right) - \dot{\hat{\sigma}}_{22,j}\left( \tau \right) \right) + \\ &+ g_{12,j} \dot{\hat{a}}_1^{\mbox{\footnotesize \dag}}\left( \tau \right) \hat{\sigma}_{13,j}\left( \tau \right) + g_{12,j} \hat{a}_1^{\mbox{\footnotesize \dag}}\left( \tau \right) \dot{\hat{\sigma}}_{13,j}\left( \tau \right) \bigg] e^{i \Delta_{2,j}\left(\tau - t \right)} . \label{eq:sigma23} \\
\end{aligned}\end{equation}
\end{widetext}

We have assumed that all atoms are initially in the ground state, so that $\hat{\sigma}_{12,j}\left( t_0 \right) = \hat{\sigma}_{23,j}\left( t_0 \right) = 0$. Further integration by parts reveals that the integral terms are second order in $\frac{1}{\Delta_{1,j}}$ and $\frac{1}{\Delta_{2,j}}$. This procedure can be repeated indefinitely to produce an infinite series in $\frac{1}{\Delta_{1,j}}$ and $\frac{1}{\Delta_{2,j}}$. We assume that $\Delta_{1,j}$ and $\Delta_{2,j}$ are large and neglect all but the first-order terms. We also assume that the cavity fields are tuned close to the two-photon resonance, so that $\Delta_{1,j}$ and $\Delta_{2,j}$ are each much greater than $\left(\Delta_{1,j} + \Delta_{2,j}\right)$. In this case it is convenient to define $\Delta_j \equiv \Delta_{1,j} \approx - \Delta_{2,j}$. These conditions for the validity of the approximation are discussed in more detail in Appendix \ref{SA1}. After the approximation we are left with
\begin{equation} \begin{aligned}
\hat{\sigma}_{12,j} & \approx \frac{1}{\Delta_{j}} \bigg[ ig_{12,j} \hat{a}_1 \left( \hat{\sigma}_{22,j} - \hat{\sigma}_{11,j} \right) - i g_{23,j} \hat{a}_2^{\mbox{\footnotesize \dag}} \hat{\sigma}_{13,j} \bigg] \\
\hat{\sigma}_{23,j} & \approx \frac{1}{\Delta_{j}} \bigg[ ig_{23,j} \hat{a}_2 \left( \hat{\sigma}_{22,j} - \hat{\sigma}_{33,j} \right) - i g_{12,j} \hat{a}_1^{\mbox{\footnotesize \dag}} \hat{\sigma}_{13,j} \bigg] . \label{eq:sigmasapprox} \\
\end{aligned} \end{equation}

We substitute these expressions into the remaining equations of motion. The resulting new equations describe the system evolution in the limit of large $\Delta_{1,j}$ and $\Delta_{2,j}$. Working backwards we can then infer an effective Hamiltonian which produces the same dynamics. We find
\begin{equation} \hat{\mathpzc{H}}_{\textnormal{eff}}=\hat{\mathpzc{H}}_{0,\textnormal{eff}} + \hat{\mathpzc{H}}_{I,\textnormal{eff}} \label{eq:Heff} \end{equation}
\begin{equation}
\hat{\mathpzc{H}}_{0,\textnormal{eff}} = \hbar \sum_{j=1}^{N} \sum_{i=1}^{3} \left( \omega_{i,j} \hat{\sigma}_{ii,j} \right) + \hbar \omega_1 \hat{a}_1^{\mbox{\footnotesize \dag}} \hat{a}_1 + \hbar \omega_2 \hat{a}_2^{\mbox{\footnotesize \dag}} \hat{a}_2 \label{eq:H0eff}
\end{equation}
\begin{equation} \begin{aligned}
\hat{\mathpzc{H}}_{I,\textnormal{eff}} &= \hbar \sum_{j=1}^{N} \bigg[ -\frac{g_{12,j}^2}{\Delta_j} \hat{a}_1^{\mbox{\footnotesize \dag}} \hat{a}_1 \left( \hat{\sigma}_{11,j} - \hat{\sigma}_{22,j} \right) - \\&- \frac{g_{23,j}^2}{\Delta_j} \hat{a}_2^{\mbox{\footnotesize \dag}} \hat{a}_2 \left( \hat{\sigma}_{33,j} - \hat{\sigma}_{22,j} \right) - \\&- \frac{g_{12,j} g_{23,j}}{\Delta_j} \left( \hat{a}_1^{\mbox{\footnotesize \dag}} \hat{a}_2^{\mbox{\footnotesize \dag}} \hat{\sigma}_{13,j} + \hat{a}_1 \hat{a}_2 \hat{\sigma}_{31,j} \right) \bigg] .  \label{eq:HIeff}
\end{aligned} \end{equation}

We note that the same mathematical procedure followed in this Section has previously been used to derive similar results \cite{Gardiner00, Gerry90, Gerry92}. Within the limits of the approximation $H_{\textnormal{eff}}$ completely determines the unitary evolution of the system. Notice that this Hamiltonian describes a system of two-level atoms which make only two-photon transitions between states $\ket{1}$ and $\ket{3}$.


\section{Classical driving field vs. single-photon driving field} \label{S3}
\subsection{Classical driving field} \label{S3.1}

We now examine the system dynamics as governed by the effective Hamiltonian of Eq's. \ref{eq:Heff}-\ref{eq:HIeff}. First we consider the case in which the $\ket{1}$ to $\ket{2}$ transition is driven with a classical field. For this case we rewrite the Hamiltonian with the substitutions $g_{12,j}\hat{a}_1\rightarrow \Omega_{12,j}/2$ and $g_{12,j}\hat{a}_1^{\mbox{\footnotesize \dag}} \rightarrow \Omega_{12,j}/2$, where $\Omega_{12,j}$ is the Rabi frequency of the classical field as seen by atom $j$. The result in the interaction picture is
\begin{equation} \begin{aligned}
\hat{\mathpzc{H}}_{\textnormal{eff}}^{\left( I \right)} &= \hbar \sum_{j=1}^{N} \bigg[ \frac{1}{2} \left( \Delta_{1,j} + \Delta_{2,j} \right) \hat{\sigma}_{z,j} + \frac{1}{2} \left( \Delta_{1,j} - \Delta_{2,j} \right) \hat{\sigma}_{22,j} - \\ & - \frac{\Omega_{12,j}^2}{4 \Delta_j} \left( \hat{\sigma}_{11,j} - \hat{\sigma}_{22,j} \right) - \frac{g_{23,j}^2}{\Delta_j} \hat{a}_2^{\mbox{\footnotesize \dag}} \hat{a}_2 \left( \hat{\sigma}_{33,j} - \hat{\sigma}_{22,j} \right) - \\&- \frac{\Omega_{12,j} g_{23,j}}{2 \Delta_j} \left( \hat{a}_2^{\mbox{\footnotesize \dag}} \hat{\sigma}_{13,j} + \hat{a}_2 \hat{\sigma}_{31,j} \right) \bigg] ,  \label{eq:CSHI'eff}
\end{aligned} \end{equation}
where we have defined $\hat{\sigma}_{z,j} \equiv \hat{\sigma}_{33,j} - \hat{\sigma}_{11,j}$. The first term in Eq. \ref{eq:CSHI'eff} represents the two-photon detuning, and includes the effects of Doppler broadening on the two-photon transition. The third and fourth terms indicate Stark shifts, which in general are different from atom to atom and will contribute to the inhomogeneous broadening.

We assume that cavity modes 1 and 2 consist of counter-propagating beams with similar wavelengths, as can be accomplished for instance in a ring cavity or whispering-gallery-mode resonator. The condition may be applied to standing-wave cavities as well, though the total interaction time must be reduced by a factor $1/2$ to account for the fact that the two modes effectively counterpropagate only half of the time  \cite{Hickman15}. In this case, and for nonrelativistic atomic speeds, the sum of the two detuning parameters can be written
\begin{equation}
\Delta_{1,j} + \Delta_{2,j} = \Delta_1 + \Delta_2 + \beta_j \left( \omega_1 - \omega_2 \right) , \label{eq:Deltas}
\end{equation}
where $\beta_j = v_j / c$ depends on the velocity of atom $j$ along the optic axis. We will assume that $\beta_j \left( \omega_1 - \omega_2 \right)$ is small and can be neglected, as is the case in many experimental systems \cite{Venkataraman13, Hickman15, Hendrickson10}. The parameters $\Delta_1$ and $\Delta_2$ represent the detuning values for atoms with $v_j = 0$, and are the same for all atoms.

For the sake of argument we now neglect the Stark shift terms as well, reducing the problem to that of a collection of N Doppler-free two-level atoms with inhomogeneous cavity coupling. The atomic states $\ket{2}$ are then completely decoupled from the rest of the system, so the term $ \left( \Delta_{1,j} - \Delta_{2,j} \right) \hat{\sigma}_{22,j}$ is constant and can be neglected as well. The Hamiltonian is
\begin{equation}\begin{gathered}
\hat{\mathpzc{H}}_{\textnormal{eff}}^{\left( I \right)} = \hbar  \sum_{j=1}^{N} \bigg[  \frac{1}{2} \left( \Delta_{1} + \Delta_{2} \right) \hat{\sigma}_{z,j} + \\ + \frac{\Omega_{12,j} g_{23,j}}{2 \Delta_j} \left( \hat{a}_2^{\mbox{\footnotesize \dag}} \hat{\sigma}_{13,j} + \hat{a}_2 \hat{\sigma}_{31,j} \right) \bigg] . \label{eq:CSHI'eff2}
\end{gathered}\end{equation}
The complete solution for an inhomogeneously coupled system such as this one can be quite complex, but when the number of excitations is small a partial solution can easily be obtained. We consider for a moment the case in which the cavity fields are tuned on resonance with the two-photon atomic transition, i.e. when $\Delta_1 + \Delta_2 = 0$. By analogy with \cite{Thompson98} we introduce the basis states
\begin{equation}
\ket{\psi_1} \equiv \ket{0_A} \ket{1_{c2}}, \ \ \ \ket{\psi_2} \equiv \ket{1_A} \ket{0_{c2}} . \label{eq:CSkets1a}
\end{equation}
The kets are defined as
\begin{equation}\begin{gathered}
\ket{0_A} \equiv \ket{1_1} \ket{1_2} ... \ket{1_N} \\
\ket{1_A} \equiv \frac{1}{g_{\textnormal{eff},0} \sqrt{N_e}} \sum_{j=1}^N g_{\textnormal{eff},j} \ket{1_1} \ket{1_2} ... \ket{3_j} ... \ket{1_N} \ , \\
\textnormal{with}\ \ \ g_{\textnormal{eff},j} = \frac{\Omega_{12,j} g_{23,j}}{2 \Delta_j} \ . \label{eq:CSkets1b}
\end{gathered}\end{equation}
Here $\ket{1_i}$ indicates the state in which atom $i$ occupies energy level $1$, etc. Then $\ket{0_A}$ represents the state with all atoms in the ground level, and $\ket{1_A}$ represents a collective excitation in which the ensemble has absorbed one photon from each cavity mode. The ket $\ket{0_{c2}}$ indicates the field state with zero excitations in mode 2, and $\ket{1_{c2}}$ the state with one excitation.

It is convenient to define
\begin{equation}
N_e \equiv \frac{1}{g_{\textnormal{eff},0}^2} \sum_{j=1}^{N} g_{\textnormal{eff},j}^2 \ ,  \label{eq:CSNe}
\end{equation}
where $N_e$ represents the effective number of atoms in the cavity given the maximum effective atom-cavity coupling rate $g_{\textnormal{eff},0}$. Diagonalizing $\hat{\mathpzc{H}}_{\textnormal{eff}}^{\left( I \right)}$ in the above basis one finds the eigenvalues
\begin{equation}
\lambda_{\pm}^{\left( CD \right)} = \pm g_{\textnormal{eff},0} \sqrt{N_e} \ . \label{eq:CSlambda}
\end{equation}

In the case of large $N$ (and hence large $N_e$) energy is exchanged between the atoms and cavity at the Rabi frequency $\Omega_{CD} = 2g_{\textnormal{eff},0}\sqrt{N_e}$, which can be much greater than both $\frac{\Omega_{12,j}^2}{4 \Delta_j}$ and $\frac{g_{23,j}^2}{\Delta_j}$. Thus our neglect of the Stark shifts in Eq. \ref{eq:CSHI'eff} is well justified and we can accept the interaction Hamiltonian of Eq. \ref{eq:CSHI'eff2} as a valid description of the system. Examining Eq. \ref{eq:CSHI'eff2} we see that when cavity mode 1 is driven with a classical field, this ensemble of $N$ Doppler-broadened three-level atoms is equivalent to a system of $N$ Doppler-free two-level atoms with inhomogeneous coupling rates $g_{\textnormal{eff},j}$.


\subsection{Single-photon driving field} \label{S3.2}

We have just seen that when a classical field drives the $\ket{1}$ to $\ket{2}$ transition, the effective Hamiltonian of our system is equivalent to the Hamiltonian of an ensemble of two-level atoms with suppressed Doppler broadening. The situation is different when this transition is driven with a single photon. In the interaction picture the Hamiltonian is
\begin{equation} \begin{aligned}
\hat{\mathpzc{H}}_{\textnormal{eff}}^{\left( I \right)} &= \hbar \sum_{j=1}^{N} \bigg[ \frac{1}{2} \left( \Delta_{1} + \Delta_{2} \right) \hat{\sigma}_{z,j} - \\ & - \frac{g_{12,j} g_{23,j}}{2 \Delta_j} \left(\hat{a}_1^{\mbox{\footnotesize \dag}} \hat{a}_2^{\mbox{\footnotesize \dag}} \hat{\sigma}_{13,j} + \hat{a}_1 \hat{a}_2 \hat{\sigma}_{31,j} \right) \bigg], \label{eq:SPHI'eff}
\end{aligned} \end{equation}
where we have neglected the Stark shift and two-photon Doppler shift terms in Eq's. \ref{eq:H0eff} and \ref{eq:HIeff} for the reasons discussed in Section \ref{S3.1}. Within our approximation of large detuning from the intermediate state, the atoms in this ensemble can undergo only two-photon transitions. Since we assume that only one photon is present in cavity mode 1, this ensemble will absorb at most one photon from mode 2 and no more. In this case the Hamiltonian is block diagonal and a complete solution can be easily obtained.
We use the basis states
\begin{equation}\begin{aligned}
\ket{\psi_{1,n_2}} &\equiv \ket{0_A} \ket{1_{c1}} \ket{\left(n_{2}\right)_{c2}} \\
\ket{\psi_{2,n_2}} &\equiv \ket{1_A} \ket{0_{c1}} \ket{\left(n_{2} - 1 \right)_{c2}} , \\ \label{eq:SPkets}
\end{aligned}\end{equation}
where now the field states of both cavity modes must be taken into account.

Considering again the case of exact two-photon resonance $\Delta_1 + \Delta_2 = 0$, and diagonalizing the Hamiltonian, we obtain
\begin{equation}
\lambda_{\pm}^{\left( SP \right)} = \pm g_{\textnormal{eff},0} \sqrt{n_2 N_e} \ , \label{eq:SPlambda}
\end{equation}
where $N_e$ is defined as in Eq. \ref{eq:CSNe} but with $g_{\textnormal{eff},j} = g_{12,j} g_{23,j} / \Delta_j$. The Rabi frequency $\Omega_{SP} = 2 g_{\textnormal{eff},0} \sqrt{n_2 N_e}$ now possesses the square-root-of-$n_2$ behavior characteristic of single-atom cavity QED systems. The reason is simple: both the single atom system and the present system have the property that they can absorb one but only one photon from the relevant cavity mode. Within our approximation, a single-photon field driving the $\ket{1}$ to $\ket{2}$ transition causes the entire ensemble to act as if it were a single atom with coupling constant $g_{\textnormal{eff},0} \sqrt{N_e}$.


\section{Discussion and conclusions}\label{S4}

We have used the well-known technique of adiabatic elimination to analyze an ensemble of 3-level room-temperature atoms in a two-mode cavity. Within our approximation, Doppler broadening can be mitigated through the use of a Doppler-free two-photon transition. This allows a warm atomic vapor in a cavity to behave like either a cold ensemble of 2-level atoms or a single cold 2-level atom, depending on the means used to drive the first atomic transition. A large number of successful experiments in quantum nonlinear optics with cavity QED have used cold atomic clouds or single trapped atoms \cite{Tiecke14, Volz14, Reiserer14,Turchette95, Kuhn02}. Our results suggest that it may be possible to perform similar experiments using warm ensembles instead. Many research groups interested in cavity QED may not have the resources to build a complex experimental system with cooling and trapping capabilities integrated into a high-finesse cavity. The ability to perform these experiments with warm vapors could allow these researchers to become productive in a field that would otherwise have been closed to them.

There are, however, disadvantages in using our model. Within the limits of the above approximation, the collective coupling strength is always smaller than that which would be experienced by a single atom optimally coupled to the same cavity, $g_{\textnormal{eff},0} \sqrt{N_e} \ll g_{23, \textnormal{max}}$. This is the case because our approximations require $\Delta_{1,j}$ and $\Delta_{2,j}$ to be large, as can be seen from Appendix \ref{SA1} Eq. \ref{eq:Conditions}. We stress, though, that this is a result of an approximation in our theoretical treatment, and need not necessarily represent a fundamental limitation of the physical system.

It has been shown that for a single three-level atom in a cavity, an effective Hamiltonian similar to that of Eq's. \ref{eq:Heff} - \ref{eq:HIeff} can be derived for arbitrary detuning values \cite{Wu97}. A generalization of that work may prove that this holds true in the case of N atoms as well. This could produce a theory allowing warm vapors to replace cold atoms in the regime of small detunings, with $g_{\textnormal{eff},0} \sqrt{N_e} \approx g_{23, \textnormal{max}}$. The solution to this problem, however, is beyond the scope of the present work.

The system described here has the advantage that with large $N$ the coupling constant $g_{\textnormal{eff},0} \sqrt{N_e}$ would be quite stable as a function of time. Experiments based on trapped atoms can suffer adverse consequences because the coupling constant varies with time, due to the random motion of the atoms inside the trap \cite{Reiserer15}. 

We note in closing that the present treatment has considered only the coherent unitary evolution of the system and has neglected dissipation and dephasing mechanisms. In particular, atomic vapor systems like those described here may suffer from  significant transit time effects, which will limit the available interaction time for coherent evolution. Decay of the field in resonator mode 1 may reduce the available interaction time as well. Dissipation and decoherence are present in all cavity QED systems however, and in many cases these additional mechanisms will not substantively worsen the decay and decoherence rates. Because of the considerable increase in experimental complexity associated with the use of cooled and trapped atoms, we expect that our results will find important practical and fundamental scientific applications.


\section*{Acknowledgments}

The author would like to acknowledge helpful discussions with J. D. Franson and T. B. Pittman. This work was supported by the NSF under grant No. 1402708.


\appendix
\section{Validity of the approximation}\label{SA1}

Here we discuss the region of validity of the above analysis. For simplicity we will assume that the number of excitations in each cavity mode is of order unity, and that $g_{12,j}$ and $g_{23,j}$ are of roughly the same size.

Our approximation consists of the neglect of the evaluations of the integrands at $\tau = t_0$ in equations \ref{eq:sigma12} - \ref{eq:sigma23}, and of the integral terms in the same equation. The former is valid whenever Eq. \ref{eq:Conditions} is satisfied. This can be seen by substituting these evaluations into the remaining equations of motion, applying the initial condition of all atoms in the ground state, and integrating. We will treat the integral terms of equations \ref{eq:sigma12} - \ref{eq:sigma23} more carefully for purposes of illustration. We denote these terms $I_{12}$ and $I_{23}$ for convenience:
\begin{widetext}
\begin{equation} \begin{aligned}
I_{12,j} &\equiv \frac{i}{\Delta_{1,j}} \int_{t_0}^{t} d\tau \bigg[ ig_{12,j} \dot{\hat{a}}_1\left( \tau \right) \left( \hat{\sigma}_{22,j}\left( \tau \right) - \hat{\sigma}_{11,j}\left( \tau \right) \right)  + ig_{12,j} \hat{a}_1\left( \tau \right) \left( \dot{\hat{\sigma}}_{22,j}\left( \tau \right) - \dot{\hat{\sigma}}_{11,j}\left( \tau \right) \right) - \\ & - i g_{23,j} \dot{\hat{a}}_2^{\mbox{\footnotesize \dag}}\left( \tau \right) \hat{\sigma}_{13,j}\left( \tau \right) - i g_{23,j} \hat{a}_2^{\mbox{\footnotesize \dag}}\left( \tau \right) \dot{\hat{\sigma}}_{13,j}\left( \tau \right) \bigg] e^{i \Delta_{1,j}\left(\tau - t \right)} \\
I_{23,j} &\equiv \frac{i}{\Delta_{2,j}} \int_{t_0}^{t} d\tau \bigg[ ig_{23,j} \dot{\hat{a}}_2\left( \tau \right) \left( \hat{\sigma}_{33,j}\left( \tau \right) - \hat{\sigma}_{22,j}\left( \tau \right) \right)  + ig_{23,j} \hat{a}_2\left( \tau \right) \left( \dot{\hat{\sigma}}_{33,j}\left( \tau \right) - \dot{\hat{\sigma}}_{22,j}\left( \tau \right) \right) + \\ & + i g_{12,j} \dot{\hat{a}}_1^{\mbox{\footnotesize \dag}}\left( \tau \right) \hat{\sigma}_{13,j}\left( \tau \right) +  i g_{12,j} \hat{a}_1^{\mbox{\footnotesize \dag}}\left( \tau \right) \dot{\hat{\sigma}}_{13,j}\left( \tau \right) \bigg] e^{i \Delta_{2,j}\left(\tau - t \right)} . \label{eq:I's} \\
\end{aligned} \end{equation}

The conditions for validity of the approximation can be made more transparent by integrating by parts again, using the equations of motion to substitute for the time derivatives of the relevant operators. This produces
\begin{equation} \begin{aligned}
&I_{12,j} = \frac{1}{\Delta_{1,j}^2} \bigg\{ g_{12,j} \hat{S}_{12} \left( \hat{\sigma}_{22,j} - \hat{\sigma}_{11,j} \right) + g_{12,j} \hat{a}_1 \bigg[ -g_{23,j} \left( \hat{a}_2 \hat{\sigma}_{32,j} - \hat{a}_2^{\mbox{\footnotesize \dag}} \hat{\sigma}_{23,j} \right) +2 g_{12,j} \left( \hat{a}_1 \hat{\sigma}_{21,j} -\hat{a}_1^{\mbox{\footnotesize \dag}} \hat{\sigma}_{12,j} \right) \bigg] + \\ & + g_{23,j} \hat{S}_{32} \hat{\sigma}_{13,j} - g_{23,j} \hat{a}_2^{\mbox{\footnotesize \dag}} \bigg[ \left( \Delta_{1,j} + \Delta_{2,j} \right) \hat{\sigma}_{13,j} - g_{12,j} \hat{a}_1 \hat{\sigma}_{23,j} + g_{23,j} \hat{a}_2 \hat{\sigma}_{12,j} \bigg] \bigg\} + \mathcal{O}\left( \frac{1}{\Delta_{1,j}^3} \right)   \\
&I_{23,j} = \frac{1}{\Delta_{2,j}^2}\bigg\{ g_{23,j} \hat{S}_{23} \left( \hat{\sigma}_{33,j} - \hat{\sigma}_{22,j} \right) + g_{23,j} \hat{a}_2 \bigg[ 2 g_{23,j} \left( \hat{a}_2 \hat{\sigma}_{32,j} - \hat{a}_2^{\mbox{\footnotesize \dag}} \hat{\sigma}_{23,j} \right) - g_{12,j} \left( \hat{a}_1 \hat{\sigma}_{21,j} - \hat{a}_1^{\mbox{\footnotesize \dag}} \hat{\sigma}_{12,j} \right) \bigg] - \\ & - g_{12,j} \hat{S}_{21} \hat{\sigma}_{13,j} + g_{12,j} \hat{a}_1^{\mbox{\footnotesize \dag}} \bigg[ \left( \Delta_{1,j} + \Delta_{2,j} \right) \hat{\sigma}_{13,j} - g_{12,j} \hat{a}_1 \hat{\sigma}_{23,j} + g_{23,j} \hat{a}_2 \hat{\sigma}_{12,j} \bigg] \bigg\} + \mathcal{O}\left(\frac{1}{\Delta_{2,j}^3} \right) , \label{eq:I'sApprox} \\
\end{aligned} \end{equation}
\end{widetext}
using $\hat{S}_{12} = \sum_{j=1}^N g_{12,j} \hat{\sigma}_{12,j}$ and $\hat{S}_{23} = \sum_{j=1}^N g_{23,j} \hat{\sigma}_{23,j}$. We have again neglected the evaluations of the integrands at $\tau=t_0$, as their effects on the integrated equations of motion are third order in $\frac{1}{\Delta_j}$. In general $I_{12,j}$ and $I_{23,j}$ can be neglected only if each contributing term can be neglected separately. As a result, each term in Eq. \ref{eq:I'sApprox} needs to be examined to determine when this is appropriate. We will restrict our discussion to the largest terms. If these are small then the remaining terms can be shown to be small as well.

First we consider the two terms
\begin{equation}\begin{aligned}
- &\frac{1}{\Delta_{1,j}^2} g_{23,j} \hat{a}_2^{\mbox{\footnotesize \dag}}\left( \Delta_{1,j} + \Delta_{2,j} \right) \hat{\sigma}_{13,j} \\
 &\frac{1}{\Delta_{2,j}^2} g_{12,j} \hat{a}_1^{\mbox{\footnotesize \dag}} \left( \Delta_{1,j} + \Delta_{2,j} \right) \hat{\sigma}_{13,j} . \label{eq:AppT1}
\end{aligned}\end{equation}
These can be neglected if $\left| \Delta_{1,j}\right| \gg \left| \Delta_{1,j}+\Delta_{2,j} \right|$ and $\left| \Delta_{2,j} \right| \gg \left| \Delta_{1,j}+\Delta_{2,j} \right|$, which will be satisfied as long as the cavity is tuned close to two-photon resonance. Terms involving $\hat{S}_{12}$, $\hat{S}_{21}$, $\hat{S}_{23}$, and $\hat{S}_{32}$ also require attention:
\begin{equation}\begin{aligned}
&\frac{1}{\Delta_{1,j}^2}g_{12,j} \hat{S}_{12} \left( \hat{\sigma}_{22,j} - \hat{\sigma}_{11,j} \right) \\
&\frac{1}{\Delta_{1,j}^2}g_{23,j} \hat{S}_{32} \hat{\sigma}_{13,j} \\
&\frac{1}{\Delta_{2,j}^2}g_{23,j} \hat{S}_{23} \left( \hat{\sigma}_{33,j} - \hat{\sigma}_{22,j} \right). \\
&- \frac{1}{\Delta_{2,j}^2} g_{12,j} \hat{S}_{21} \hat{\sigma}_{13,j} \\ \label{eq:AppT2}
\end{aligned}\end{equation}
These can be neglected if $\left| \Delta_{1,j} \right|$ and $\left| \Delta_{2,j} \right|$ are much greater than $g_0^{\left( 1 \right)} \sqrt{N_e^{\left( 1 \right)}}$ and $g_0^{\left( 2 \right)} \sqrt{N_e^{\left( 2 \right)}}$, with
\begin{equation}\begin{aligned}
N_e^{\left( 1 \right)} = \frac{1}{{g_0^{\left( 1 \right)}}^2} \sum_{j=1}^{N} g_{12,j}^2 \\
N_e^{\left( 2 \right)} = \frac{1}{{g_0^{\left( 2 \right)}}^2} \sum_{j=1}^{N} g_{23,j}^2 . \label{eq:AppNe1}
\end{aligned}\end{equation}
If these conditions are satisfied it can be shown that all terms in Eq. \ref{eq:I'sApprox} can be neglected. We assume that in this case all third- and higher-order terms are negligible as well. Sufficient conditions for the validity of our approximation can then be written as
\begin{equation}\begin{gathered}
\left| \Delta_{1,j}\right| \gg \left| \Delta_{1,j}+\Delta_{2,j} \right|, \ \ \ \left| \Delta_{2,j} \right| \gg \left| \Delta_{1,j}+\Delta_{2,j} \right| \\
\left| \Delta_{j} \right| \gg g_0^{\left( 1 \right)} \sqrt{N_e^{\left( 1 \right)}}, \ \ \  \left| \Delta_{j} \right| \gg g_0^{\left( 2 \right)} \sqrt{N_e^{\left( 2 \right)}}. \label{eq:Conditions}
\end{gathered}\end{equation} \\


\section{Lambda-type transition diagram}\label{SA2}

\begin{figure}
  \includegraphics[width=.7\columnwidth]{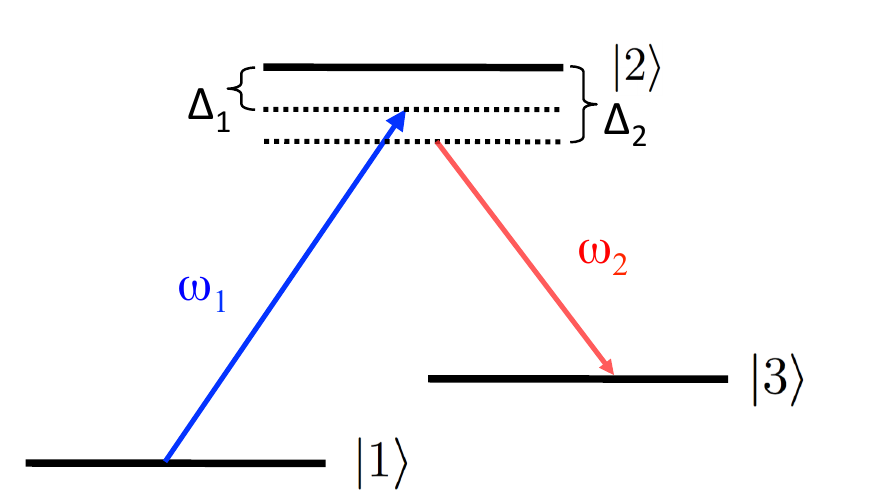}
  \caption{Energy level diagram for a three-level atom in the $\Lambda$-type configuration.}\label{fig:ElevelsLambda}
\end{figure}

A result nearly identical to that of Eq's. \ref{eq:Heff}-\ref{eq:HIeff} can be obtained when the atomic levels follow a $\Lambda$-type diagram as in Fig. \ref{fig:ElevelsLambda}. In this case the effective Hamiltonian is:
\begin{equation} \hat{\mathpzc{H}}_{eff}^{\left( \lambda \right)}=\hat{\mathpzc{H}}_{0,eff}^{\left( \lambda \right)} + \hat{\mathpzc{H}}_{I,eff}^{\left( \lambda \right)} \label{eq:HeffL} \end{equation}
\begin{equation}
\hat{\mathpzc{H}}_{0,eff}^{\left( \lambda \right)} = \hbar \sum_{j=1}^{N} \bigg[ \omega_{1,j} \hat{\sigma}_{11,j} + \omega_{3,j} \hat{\sigma}_{33,j} \bigg] + \hbar \omega_1 \hat{a}_1^{\mbox{\footnotesize \dag}} \hat{a}_1 + \hbar \omega_2 \hat{a}_2^{\mbox{\footnotesize \dag}} \hat{a}_2 \label{eq:H0effL}
\end{equation}
\begin{equation} \begin{aligned}
\hat{\mathpzc{H}}_{I,eff}^{\left( \lambda \right)}&= \hbar \sum_{j=1}^{N} \bigg[ -\frac{g_{12,j}^2}{\Delta_j} \hat{a}_1^{\mbox{\footnotesize \dag}} \hat{a}_1 \hat{\sigma}_{11,j} - \frac{g_{23,j}^2}{\Delta_j} \hat{a}_2^{\mbox{\footnotesize \dag}} \hat{a}_2 \hat{\sigma}_{33,j} - \\&- \frac{g_{12,j} g_{23,j}}{\Delta_j} \left( \hat{a}_1^{\mbox{\footnotesize \dag}} \hat{a}_2 \hat{\sigma}_{13,j} + \hat{a}_1 \hat{a}_2^{\mbox{\footnotesize \dag}} \hat{\sigma}_{31,j} \right) \bigg] ,  \label{eq:HIeffL}
\end{aligned} \end{equation}

with $\Delta_{j} \equiv \Delta_{1,j} \approx \Delta_{2,j}$. Factors of $\hat{\sigma}_{22,j}$ have been neglected here since the populations of state $\ket{2}$ will be small compared to those of states $\ket{1}$ and $\ket{3}$.

\end{document}